\documentclass[]{aastex631}
\usepackage{makecell}
\usepackage{amsmath}
\usepackage{cases}
\usepackage{multirow}

\begin{document}

\title{The GeV $\gamma$-ray emission from the composite SNR CTB 87}

\author[0009-0004-5450-8237]{Yuliang Xin}
\affiliation{School of Physical Science and Technology, Southwest Jiaotong University, Chengdu, Sichuan, 610031, People’s Republic of China;  \href{mailto:ylxin@swjtu.edu.cn}{ylxin@swjtu.edu.cn}}
\author{Jian Tang}
\affiliation{Department of Physics and Electronic Science, Aba Teachers University, Wenchuan, Sichuan, 623002, People’s Republic of China;  \href{huzida@126.com}{huzida@126.com}}
\author{Weixiong Ding}
\author{Xi Liu}
\affiliation{School of Physical Science and Technology, Southwest Jiaotong University, Chengdu, Sichuan, 610031, People’s Republic of China;  \href{mailto:ylxin@swjtu.edu.cn}{ylxin@swjtu.edu.cn}}
\author{Yunfeng Zhang}
\affiliation{School of Physics and Engineering Technology, Chengdu Normal University, Chengdu, Sichuan, 611130, People’s Republic of China}
\author{Xiaolei Guo}
\affiliation{School of Physical Science and Technology, Southwest Jiaotong University, Chengdu, Sichuan, 610031, People’s Republic of China;  \href{mailto:ylxin@swjtu.edu.cn}{ylxin@swjtu.edu.cn}}



\begin{abstract}
We report the GeV $\gamma$-ray emission around the composite supernova remnant (SNR) CTB 87 with more than 16 yrs PASS 8 data recorded by the Fermi Large Area Telescope. Two separate point sources with the different GeV spectra are identified in this region: one has a soft $\gamma$-ray spectrum, likely due to interactions between the SNR shock and molecular clouds (MCs); and another source with a hard GeV $\gamma$-ray spectrum aligns with the TeV spectrum of VER J2016+371, suggesting it as the GeV counterpart. Considering the observations of CTB 87 in the radio and X-ray bands, VER J2016+371 is proposed to originate from the pulsar wind nebula (PWN) associated with PSR J2016+3711. A leptonic model with a broken power-law electron distribution could explain the multi-wavelength data of VER J2016+371, with fitted parameters matching typical $\gamma$-ray PWNe. Deeper searching for the SNR shock of CTB 87 in other bands and the future TeV observations by LHAASO and CTA are crucial to reveal the nature of CTB 87.
\end{abstract}



\section{Introduction} 
\label{sec:intro}
SNRs are widely believed to be the dominant accelerators of Galactic cosmic rays \citep[CRs;][]{1934PNAS...20..259B, 2022RvMPP...6...19L}. PWNe, one of the most important components of SNRs, are dynamic and energetic structures powered by the rotational energy of pulsars. These nebulae are formed by the interaction of magnetized relativistic particles with the surrounding medium \citep{2006ARA&A..44...17G}. The accelerated particles in SNRs and PWNe can produce a broad spectrum of electromagnetic radiation, ranging from radio to $\gamma$-ray frequencies. 
The detection of radio and nonthermal X-ray emissions provides clear evidence of electron acceleration \citep{1995Natur.378..255K, 2006ApJ...641..427C}. 
$\gamma$-ray emissions can arise from hadronic interactions, where $\pi^0$ decays into two $\gamma$-ray photons, through inverse Compton (IC) scattering, or via nonthermal bremsstrahlung radiation from high-energy electrons. 
$\gamma$-ray emissions from SNRs and PWNe have been extensively observed by space telescopes, particularly the Large Area Telescope (LAT) aboard the {\em Fermi} satellite, and ground-based Cherenkov telescopes such as H.E.S.S. \citep{2004APh....22..109A}, VERITAS \citep{2006APh....25..391H}, MAGIC \citep{2016APh....72...61A}, HAWC \citep{2013APh....50...26A}, and LHAASO \citep{2024ApJS..271...25C}. 
Detailed studies of these $\gamma$-ray SNRs and PWNe are crucial for understanding the acceleration and radiation processes of particles, and for further elucidating the origin of CRs.
Here we report on the GeV $\gamma$-ray emission from the SNR CTB 87 associated with the TeV $\gamma$-ray source VER J2016+371.

SNR CTB 87, also known as G74.9+1.2, is classified as a filled-centered SNR, which means it is filled with synchrotron emission from a PWN rather than a shell-like structure.
The radio observations exhibit a kidney-shaped morphology of CTB 87 with the radio size of about 8$^{'}$ $\times$ 6$^{'}$ \citep{2009BASI...37...45G}. 
The radio spectral index of CTB 87 is $\alpha$ $\approx$ $-$0.29 up to a frequency of about 10 GHz \citep{2006A&A...457.1081K}. 
And a remarkable spectral steepening above 11 GHz was firstly reported by \citet{1987A&AS...69..533M} with the spectral index changes to $\alpha$ $\approx$ $-$1.08 based on Effelsberg 32-GHz observations,
which was supported by the observations from IRAM 30 m telescope at 84 GHz \citep{1989ApJ...338..171S} and Arcminute Microkelvin Imager (AMI) at frequencies from 14 to 18 GHz \citep{2009MNRAS.396..365H}. 
Using the radio observations between 4.75 and 32 GHz from the Effelsberg 100-m Radio Telescope and the archived low-frequency observations at 1420 and 408 MHz from the Canadian Galactic Plane Survey (CGPS), \citet{2020MNRAS.496..723K} distinguished two separate emission components of CTB 87: a compact, kidney-shaped component characterized by a steeper spectrum, and a larger, diffuse, spherical component that is centrally peaked.
The compact component with an size of about 7.8$^{'}$ $\times$ 4.8$^{'}$ is suggested to be a relic PWN, and the diffuse component with an size of about 17$^{'}$ in diameter represents the undisturbed part of the PWN expanding inside a cavity or stellar wind bubble \citep{2020MNRAS.496..723K}.
The existence of two components for CTB 87 was subsequently confirmed by \citet{2022A&A...668A..39R} with the new $\lambda$2 cm observations by Effelsberg 100 m telescope.
The distance to CTB 87 has been a subject of debate, with earlier estimates suggesting a distance of 12 kpc based on HI absorption measurements \citep{1989MNRAS.237..555G, 1997A&A...317..212W}. 
While using the extinction-distance relation, an updated distance of 6.1 $\pm$ 0.9 kpc was established \citep{2003ApJ...588..852K}.

In the X-ray band, \citet{2013ApJ...774...33M} presented the first detailed X-ray study of CTB 87 with {\em Chandra}, and found an extended diffuse nebula around a point source, CXOU J201609.2+371110, which is suggested to be the putative pulsar powering CTB 87.
The diffuse nebula follows a power-law spectrum with an index of $\sim$1.68.
{\em XMM}-Newton confirmed the detection of nebula, and also found an overall steepening of the photon index away from the putative pulsar \citep{2020MNRAS.491.3013G}.
The non-detection of thermal X-ray emission from SNR by {\em XMM}-Newton supports the scenario for a $\sim$20-kyr-old relic PWN expanding into a stellar wind-blown bubble.
The compact nature of CXOU J201609.2+371110 was confirmed by the observations of the Five-hundred-meter Aperture Spherical radio Telescope (FAST), which detected the radio pulse profile and revealed it to be a pulsar \citep[PSR J2016+3711;][]{2024MNRAS.528.6761L}.
The characteristic age and spin-down luminosity of PSR J2016+3711 are 11.1 kyr and 2.2 $\times$10$^{\rm 37}$ erg s$^{\rm -1}$, respectively.
%{\em Chandra} observations, find the peak of the X-ray emission is clearly offset from the peak of the radio emission and located at the southeastern edge of the radio nebula; -- which can be attributed to the neutron star's motion,  -- age of $\sim$ 5-28 kyr,

The $\gamma$-ray emission from CTB 87 was firstly detected by VERITAS with a point-like morphology, named as VER J2016+371 \citep{2014ApJ...788...78A}.
The TeV $\gamma$-ray spectrum of VER J2016+371 in the energy range of 0.68 - 14.7 TeV is a power-law distribution with an index of 2.1 $\pm$ 0.8$_{\rm stat}$ $\pm$ 0.4$_{\rm sys}$ \citep{2018ApJ...861..134A}. 
\citet{2016MNRAS.460.3563S} analyzed the {\em Fermi}-LAT data with 3FGL catalog and found a point source spatially consistent with CTB 87.
The GeV $\gamma$-ray spectrum described by a log-parabola function is soft with no significant emission above 10 GeV.
And they suggested that the multi-wavelength data of CTB 87 from radio to TeV bands could be interpreted by a Maxwellian distribution of electrons along with a broken power-law distribution of electrons in low magnetic fields \citep{2016MNRAS.460.3563S}.
CTB 87 exhibits significant interactions with the nearby MCs, which has been observed through millimeter CO-line observations \citep{2003ApJ...588..852K, 2018ApJ...859..173L, 2023ApJS..268...61Z}.
\citet{2018ApJ...859..173L} detected the interaction between SNR and the MCs based on the asymmetric broad profiles of $^{\rm 12}$CO lines at -58 km s$^{\rm -1}$, particularly at the eastern and southwestern edges of the radio emission.
They suggested that the kidney-shaped radio emission of CTB 87 represents the relic of the part of the blast wave that has been driven into the MC complex, which makes CTB 87 to be a composite system.
The SNR–MCs interaction with the hadronic process was thought to contribute to the $\gamma$-ray emissions from the CTB 87 region \citep{2016MNRAS.460.3563S, 2018ApJ...859..173L}.

In this study, we present the re-analyzed results of the GeV $\gamma$-ray emission around CTB 87, utilizing the PASS 8 data collected by {\em Fermi}-LAT. 
The method and results of our data analysis, encompassing both spatial and spectral analysis, are detailed in Section 2. 
The potential origins of the different $\gamma$-ray emission components are discussed in Section 3, with a summary provided in Section 4.

\section{Fermi-LAT Data and Results}
\label{sec:fermi-data}

\subsection{Data Reduction}
\label{data reduction}
In the following analysis, we select the latest Pass 8 version of {\em Fermi}-LAT data recorded from August 4, 2008 (Mission Elapsed Time 239557418) to September 4, 2024 (Mission Elapsed Time 747100805) with ``Source'' event class (evclass = 128 \& evtype = 3) to analyze the $\gamma$-ray emission around CTB 87.
The energy range of data is selected to be from 1 GeV to 1 TeV by considering the improved point-spread function (PSF) at higher energies.
To reduce the contamination from Earth Limb, the events whose zenith angles larger than $90^\circ$ are excluded.
The region of interest (ROI) is a $20^\circ \times 20^\circ$ square region centered at the position of CTB 87 (R.A. = $304^{\circ}\!.057$, decl. = $37^{\circ}\!.213$).
The data are analyzed using the standard {\it Fermi ScienceTools} \footnote {http://fermi.gsfc.nasa.gov/ssc/data/analysis/software/} 
with the instrumental response function (IRF) of ``P8R3{\_}SOURCE{\_}V3'' and the binned likelihood analysis method.
The Galactic and isotropic diffuse background models used here are {\tt gll\_iem\_v07.fits} and {\tt iso\_P8R3\_SOURCE\_V3\_v1.txt}
\footnote {http://fermi.gsfc.nasa.gov/ssc/data/access/lat/BackgroundModels.html}, respectively.
All sources in the incremental version of the fourth {\em Fermi}-LAT source catalog \citep[4FGL-DR4;][]{2020ApJS..247...33A, 2023arXiv230712546B}, within a radius of $20^\circ$ from the ROI center and the two components of the diffuse background, are included in the source model, which is generated by the user-contributed software 
{\tt make4FGLxml.py}\footnote{http://fermi.gsfc.nasa.gov/ssc/data/analysis/user/}.
During the likelihood analysis, the normalizations and the spectral parameters of all sources within $5^{\circ}$ to the center of ROI, together with the normalizations of the two components of the diffuse background, are set to be free.

\subsection{Spatial Analysis}
\label{Spatial}

\begin{figure*}[!htb]
	\centering
    \includegraphics[width=0.8\textwidth]{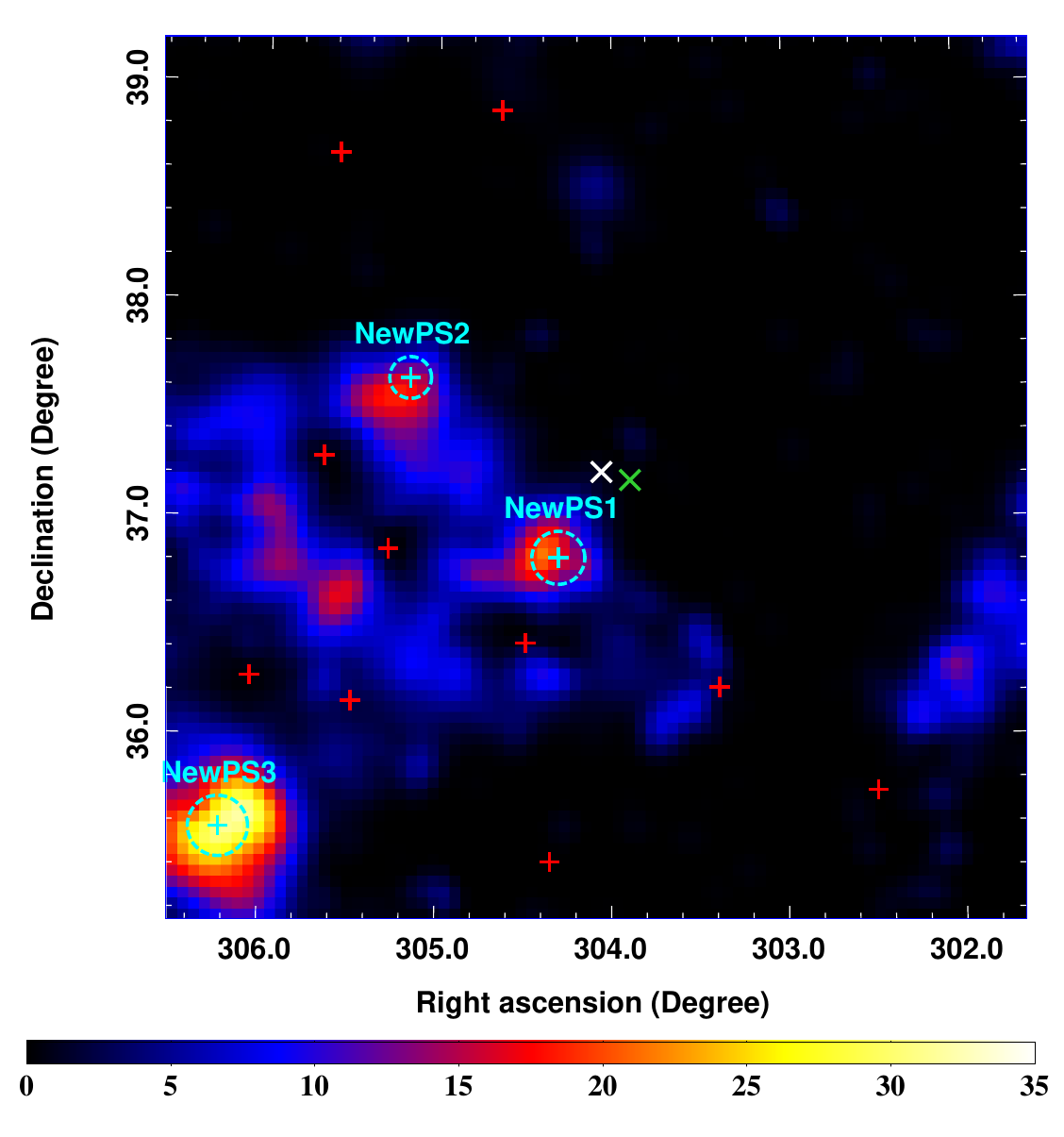}
	\caption{$4^{\circ}\!.0$ $\times$ $4^{\circ}\!.0$ TS map in the energy range of 1 GeV - 1 TeV by subtracting the emission from 4FGL-DR4 sources and the diffuse backgrounds. The red pluses mark the sources in the 4FGL-DR4 catalog, and 4FGL J2016.2+3712 and 4FGL J2015.5+3710 are shown as the green and white crosses, respectively. The best-fit positions of three newly added point sources are indicated as the cyan pluses, with the 1$\sigma$ uncertainty of each source marked by a cyan dashed circle.}
	\label{fig:tsmap-4fgl-resi}
\end{figure*}

\begin{table}[!htb]
\centering
\caption {Coordinates of the three newly added point sources}
\begin{tabular}{cccc}
\hline 
Name & R.A., decl.  & 1$\sigma$ uncertainty \\
\hline
NewPS1  & $304^{\circ}\!.3038 \pm 0^{\circ}\!.0803$, $36^{\circ}\!.8189 \pm 0^{\circ}\!.0803$ & 0$^\circ$.1217\\
NewPS2  & $305^{\circ}\!.1620 \pm 0^{\circ}\!.0806$, $37^{\circ}\!.6419 \pm 0^{\circ}\!.0498$ & 0$^\circ$.0961\\
NewPS3  & $306^{\circ}\!.2242 \pm 0^{\circ}\!.0970$, $35^{\circ}\!.5721 \pm 0^{\circ}\!.0992$ & 0$^\circ$.1382\\
\hline
\end{tabular}
\label{table:newpts}
\end{table}

\begin{table}[!htb]
\centering
\caption {Spatial Analysis in the different energy ranges}
\begin{tabular}{ccccccccc}
\hline 
Energy Range& Source  &  R.A., decl. (1$\sigma$ uncertainty) & Index  & TS Value & -log(Likelihood)\\
%&         & (1$\sigma$ uncertainty) & $\Gamma$ or $\alpha$/$\beta$ &Value &\\
%Range  &         & (1$\sigma$ uncertainty) & $\Gamma$ or $\alpha$/$\beta$ &Value &\\
\hline
\multirow{2}{*}{1$-$30 GeV} &  J2016.2+3712  & $304^{\circ}\!.0557 \pm 0^{\circ}\!.0150$, $37^{\circ}\!.2014 \pm 0^{\circ}\!.0120$ (0$^\circ$.0203)  &  2.574$\pm$0.138  & 270.881   & -1623144.16\\ 
%&  J2015.5+3710  & $303^{\circ}\!.8798$, $37^{\circ}\!.1711$  &  2.429$\pm$0.061/0.145$\pm$0.046  & 2671.97  & --\\ 
\cline{2-6}
& PsA  & $304^{\circ}\!.0557$, $37^{\circ}\!.2014$  &  2.473$\pm$0.185   & 262.83  & \multirow{2}{*}{-1623144.98} \\ 
 & PsB  & $304^{\circ}\!.0490$, $37^{\circ}\!.2161$  &  --  & 1.02   &  \\
\hline
 \multirow{2}{*}{30GeV$-$1TeV} & J2016.2+3712  & $304^{\circ}\!.0490 \pm 0^{\circ}\!.0173$, $37^{\circ}\!.2161 \pm 0^{\circ}\!.0171$ (0$^\circ$.0261)  &  1.394$\pm$0.343  & 29.16   & 25092.27 \\ 
%&  J2015.5+3710  & $303^{\circ}\!.8798$, $37^{\circ}\!.1711$  &  2.880$\pm$4.011/0.120$\pm$0.599  & 5.59   & --\\ 
\cline{2-6}
& PsA  & $304^{\circ}\!.0557$, $37^{\circ}\!.2014$  &  --   & 0.0   & \multirow{2}{*}{25092.24} \\ 
 & PsB  & $304^{\circ}\!.0490$, $37^{\circ}\!.2161$  &  1.396$\pm$0.357  & 28.81   &  \\ 
\hline
1GeV$-$1 TeV & \multirow{2}{*}{J2016.2+3712}  &  \multirow{2}{*}{$304^{\circ}\!.0755 \pm 0^{\circ}\!.0096$, $37^{\circ}\!.2021 \pm 0^{\circ}\!.0078$ (0$^\circ$.0129)}  &   \multirow{2}{*}{2.467$\pm$0.111}  &  \multirow{2}{*}{237.19}   &  \multirow{2}{*}{-1615349.19} \\ 
 (Model 1) &&&&&\\
\cline{2-6}
%&  J2015.5+3710  & $303^{\circ}\!.8798 \pm 0^{\circ}\!.0047$, $37^{\circ}\!.1711 \pm 0^{\circ}\!.0047$ (0$^\circ$.0071)  &  2.467$\pm$0.049/0.175$\pm$0.037  & 3031.34   & --\\ 
%\hline
\multirow{2}{*}{(Model 2)} & PsA  & $304^{\circ}\!.0557$, $37^{\circ}\!.2014$  &  2.902$\pm$0.109   & 230.13   & \multirow{2}{*}{-1615355.97} \\ 
 & PsB  & $304^{\circ}\!.0490$, $37^{\circ}\!.2161$  &  1.689$\pm$0.232  & 29.09   &  \\ 
% &  J2015.5+3710  & $303^{\circ}\!.8798$, $37^{\circ}\!.1711$  &  2.373$\pm$0.057/0.225$\pm$0.039  & 2623.38   & --\\ 
\hline
\end{tabular}
\label{table:spatial}
\end{table} 

\begin{figure*}[!htb]
	\centering
    \includegraphics[width=0.48\textwidth]{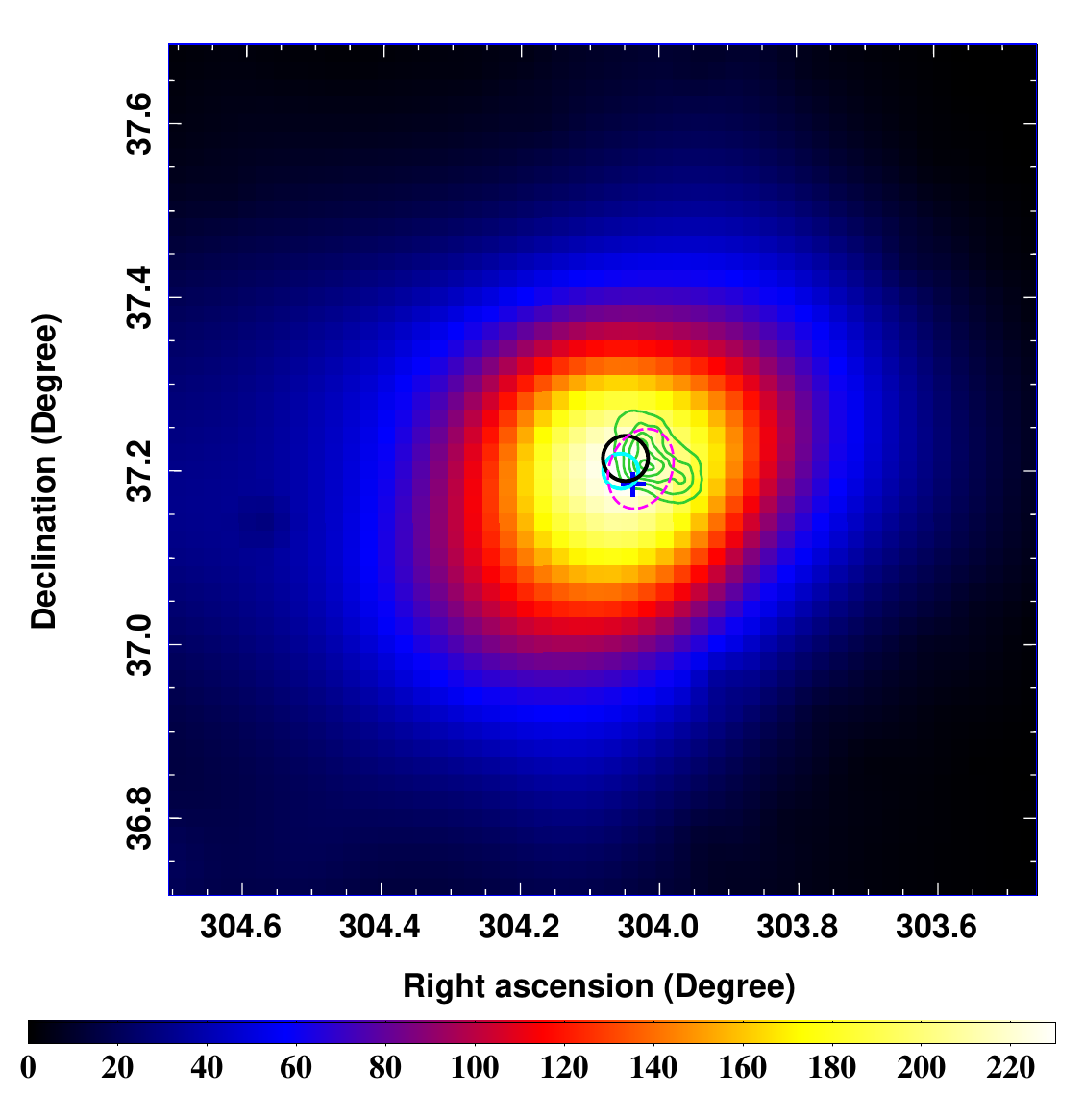}
    \includegraphics[width=0.48\textwidth]{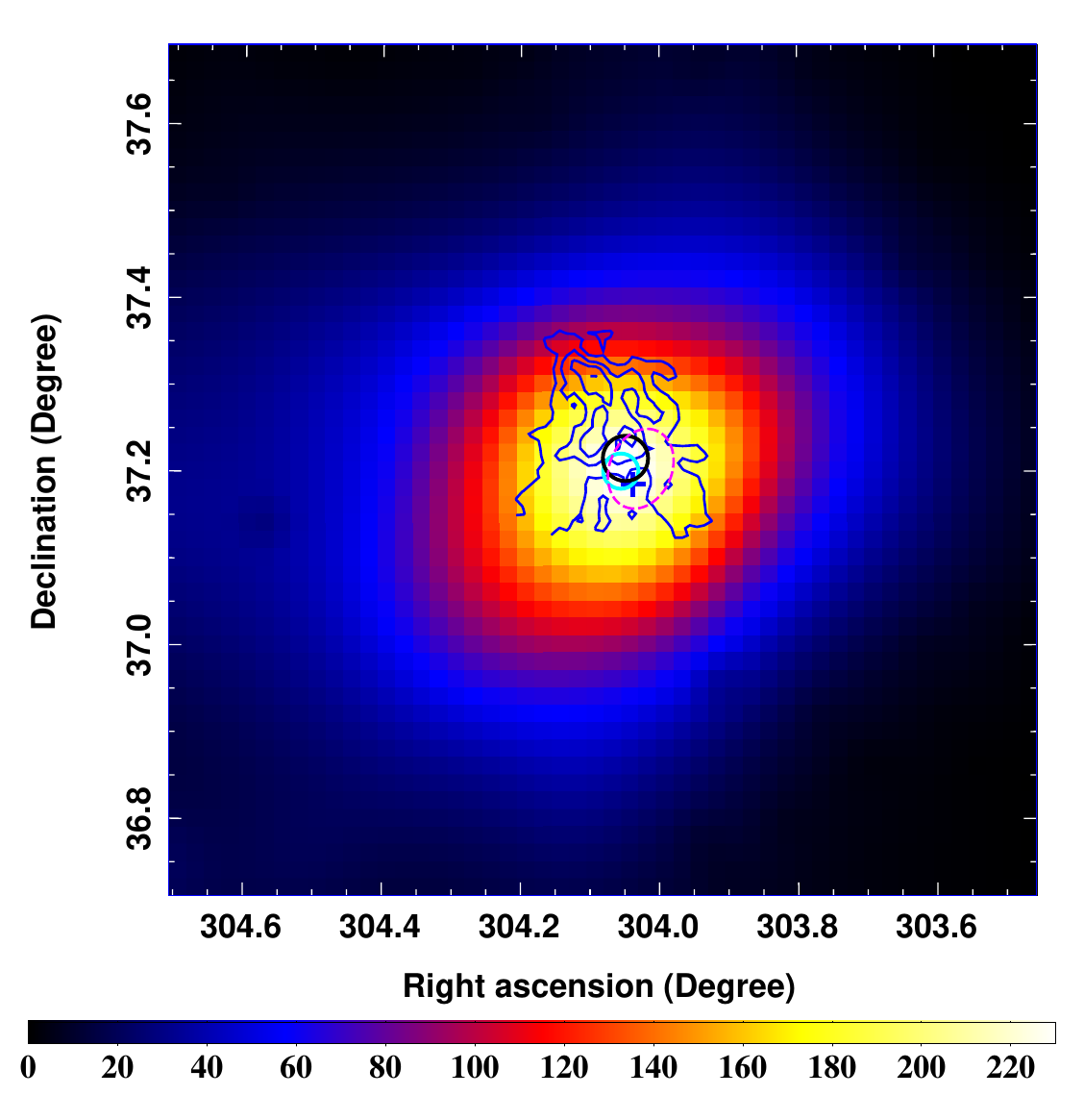}\\
    \includegraphics[width=0.48\textwidth]{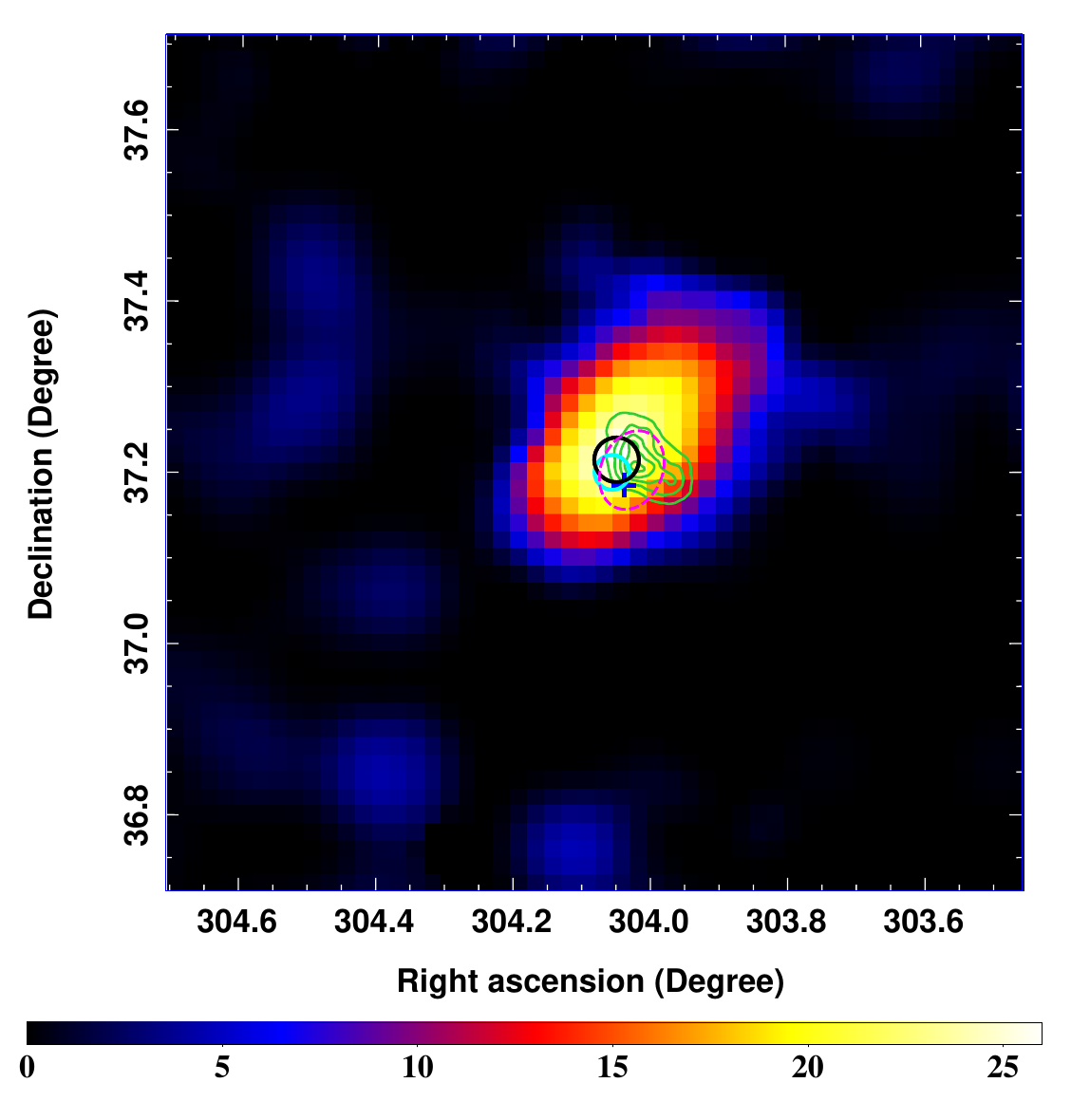}
    \includegraphics[width=0.48\textwidth]{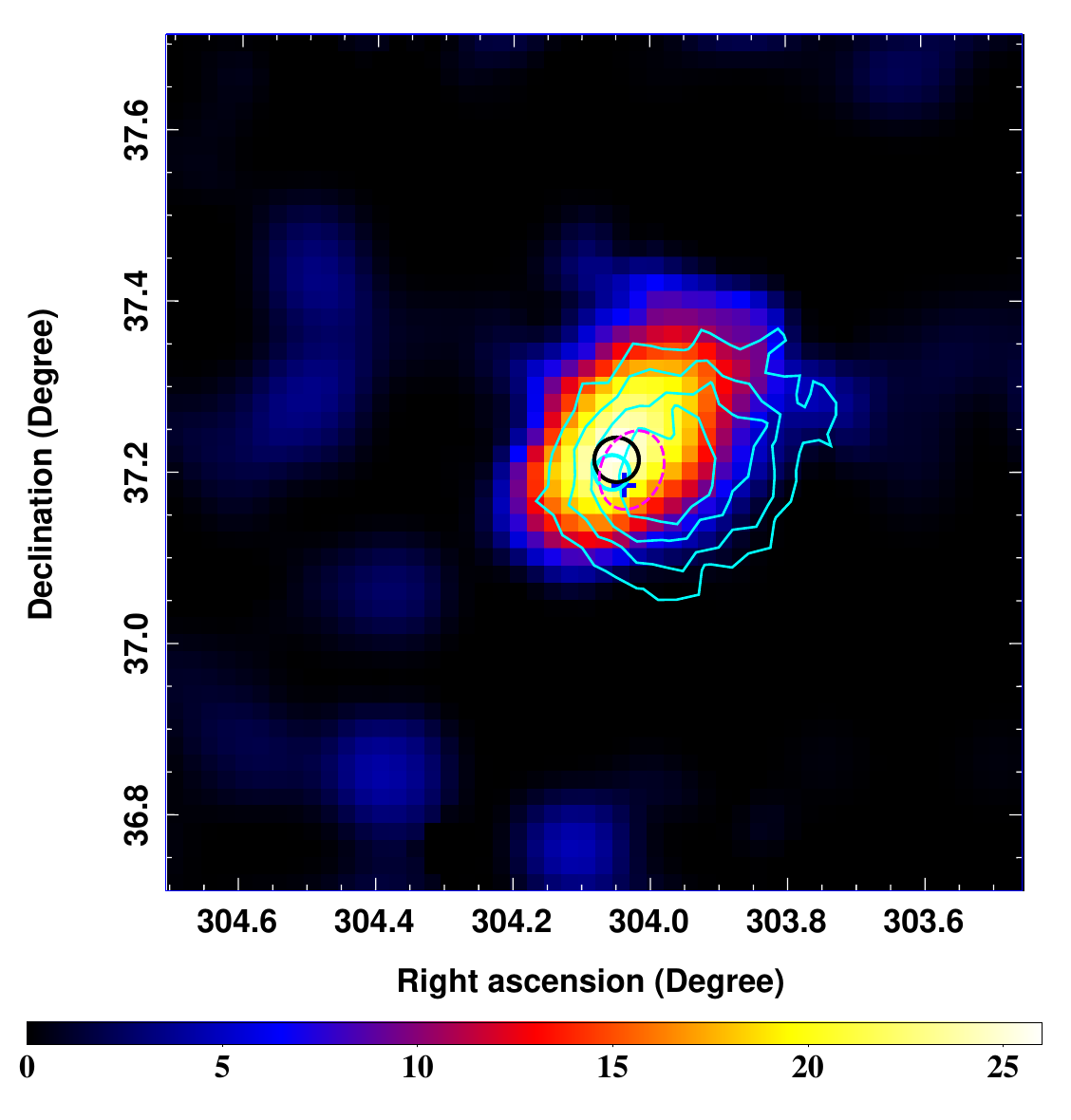}
	\caption{$1^{\circ}\!.0$ $\times$ $1^{\circ}\!.0$ TS maps in the energy range of 1 - 30 GeV (top) and 30 GeV - 1 TeV (bottom). The cyan (black) solid circle indicates the best-fit position of PsA (PsB) in the energy range of 1 - 30 GeV (30 GeV - 1 TeV) with a 68\% uncertainty radius. The X-ray emission region of CTB 87 observed by {\em Chandra} is shown as the magenta dashed ellipse \citep{2013ApJ...774...33M}. The blue plus marks the position of PSR J2016+3711 \citep{2024MNRAS.528.6761L}. The green contours in the left panels represent the radio emission of CTB 87 at 1.4 GHz \citep{2020MNRAS.496..723K}. The blue contours in the top-right panel represent the molecular clouds around CTB 87 traced by $^{\rm 12}$CO (J=1-0) intensity in velocity interval from -60 to -54 km s$^{\rm -1}$ \citep{2018ApJ...859..173L}. The TeV $\gamma$-ray emission from VER J2016+371 are shown as  the cyan contours in the bottom-right panel \citep{2018ApJ...861..134A}.}
	\label{fig:tsmaps-multiwavelength}
\end{figure*}

In the 4FGL-DR4 catalog, the $\gamma$-ray point source, 4FGL J2016.2+3712, is suggested to be the GeV counterpart of CTB 87. 
And another $\gamma$-ray point source, 4FGL J2015.5+3710, lies $0.14^\circ$ from the catalog location of 4FGL J2016.2+3712 and is associated with the blazar QSO J2015+371.
The spectra of 4FGL J2016.2+3712 and 4FGL J2015.5+3710 are adopted to be a power-law and log-parabola models, as recommended in 4FGL-DR4 catalog.
Using the data from 1 GeV to 1 TeV, we first create a Test Statistic (TS) map with {\tt gttsmap} by subtracting the emission from the sources and backgrounds in the best-fit model including 4FGL J2016.2+3712 and 4FGL J2015.5+3710, which is shown in Fig \ref{fig:tsmap-4fgl-resi}.
The TS map shows that there are extra sources beyond the 4FGL-DR4 catalog, and we mark three bright new sources with cyan pluses in this plot.
The accurate positions of them as point sources are obtained using {\tt Fermipy}, a {\tt PYTHON} package that automates analyses with {\it ScienceTools} \citep{2017ICRC...35..824W}, which are listed in Table \ref{table:newpts}.
Then we add the three additional point sources with power-law spectra in the model. 
The positions of 4FGL J2016.2+3712 and 4FGL J2015.5+3710 are also refitted with {\tt Fermipy}, which are R.A. = $304^{\circ}\!.755 \pm 0^{\circ}\!.0096$, decl. = $37^{\circ}\!.2021 \pm 0^{\circ}\!.0078$ with 1$\sigma$ uncertainty of
0$^\circ$.0129, and R.A. = $303^{\circ}\!.8798 \pm 0^{\circ}\!.0047$, decl. = $37^{\circ}\!.1711 \pm 0^{\circ}\!.0047$ with 1$\sigma$ uncertainty of 0$^\circ$.071, respectively.
With the updated spatial model (Model 1), we refit the data from 1 GeV to 1 TeV and get the GeV power-law spectrum of 4FGL J2016.2+3712 with an index of 2.467$\pm$0.111.

Then we do the same likelihood fitting in two energy ranges: 1$–$30 GeV and 30 GeV$–$1 TeV (hereafter referred to as the low and high energy ranges, respectively), to research into the energy-dependent behavior of the $\gamma$-ray emission from CTB 87.
For the models of the two energy ranges, the position of 4FGL J2015.5+3710 is fixed to be the best-fit coordinate in 1 GeV$-$1 TeV, and we refit that of 4FGL J2016.2+3712 in the low and high energy ranges, respectively.
The best-fit positions of 4FGL J2016.2+3712 in the low and high ranges are R.A. = $304^{\circ}\!.0557 \pm 0^{\circ}\!.0150$, decl. = $37^{\circ}\!.2014 \pm 0^{\circ}\!.0120$ with 1$\sigma$ uncertainty of 0$^\circ$.0203, 
and R.A. = $304^{\circ}\!.0490 \pm 0^{\circ}\!.0173$, decl. = $37^{\circ}\!.2161 \pm 0^{\circ}\!.0171$ with 1$\sigma$ uncertainty of 0$^\circ$.0261.
The centroids of the $\gamma$-ray emission in the two energy ranges are only 0$^\circ$.016 apart, which is too close to clearly resolved by considering the uncertainties of the positions, as shown in Figure \ref{fig:tsmaps-multiwavelength}.
However, the fitting spectrum of 4FGL J2016.2+3712 in the high energy range is very hard with an index of 1.394$\pm$0.343, which is much different from the result in the low energy range with an index of 2.574$\pm$0.138.
The different spectra suggest two different components in the low and high bands, which are labeled as ``PsA'' for the 1$-$30 GeV source and ``PsB'' for the 30 GeV $-$ 1 TeV source hereafter.
We then delete 4FGL J2016.2+3712 and add PsA and PsB as point sources in the model file. 
We assume that their spectra are still power-laws, and then re-fit the data again.
The fitting results in the low energy range show that the emission is dominated by PsA with a TS value of 262.83, while PsB is very weak with a TS value of 1.02.
Contrary to that in the low band, the emission from PsA in the high band is non-detectable, while PsB has significant GeV $\gamma$-ray emission with a TS value of 28.81.
%extension:
We also test the extension significance of PsA and PsB in the low and high energy bands, respectively.
Here, we treat PsB as a point source and use the two-dimensional (2D) Gaussian models centered at the best-fit position with different values of $\sigma$ as the spatial templates for PsA and re-do the fittings in the low energy range.
And for the high energy range, PsA is treated as a point source and the 2D Gaussian models are applied to PsB.
By comparing the fitting likelihood values, there are no significant extension for both PsA and PsB, and we still treat them as point sources.
Further we adopt the updated model that 4FGL J2016.2+3712 is replaced by PsA and PsB (Model 2) in the energy range of 1 GeV $-$ 1 TeV.
The overall maximum likelihood values of Model 1 and 2 are listed in Table \ref{table:spatial}.
We adopt the Akaike information criterion \citep[AIC;][]{1974ITAC...19..716A}\footnote{AIC = -2ln$\mathcal{L}$+2{\em k}, where $\mathcal{L}$ is the value of maximum likelihood and {\em k} is the parameter numbers of model.} to compare the two models by calculating $\Delta$AIC.
The value of $\Delta$AIC = AIC$_{\rm Model 2}$ - AIC$_{\rm Model 1}$ $\approx$ -9.6 suggests that the model including two points sources in this region is favored.
And we adopt the two-source model in the following spectral analysis.
The TS values of PsA and PsB are fitted to be 230.13 and 29.09 in the energy range of 1 GeV $-$ 1 TeV, corresponding to the significance level of $\sim$15.0$\sigma$ and $\sim$5.0$\sigma$ with two degrees of freedom.

\subsection{Spectral Analysis}
\label{Spectral}

\begin{figure*}[!htb]
	\centering
    \includegraphics[width=0.8\textwidth]{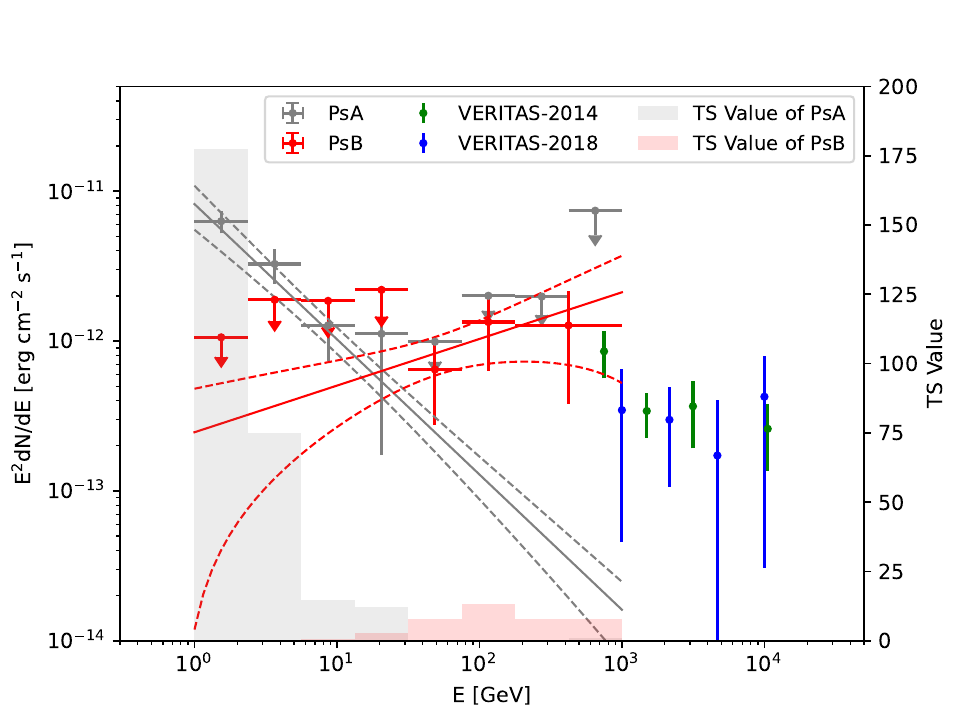}
	\caption{SEDs of PsA (gray dots) and PsB (red dots) in the energy range of 1 GeV $-$ 1 TeV with the corresponding colored histogram shown as the TS value for each energy bin. The arrows indicate the 95\% upper limits for the energy bin with TS value smaller than 5.0. The solid and dashed lines show the global best-fit power-law spectrum and its 1$\sigma$ statistic error for PsA and PsB in the energy range of 1 GeV $-$ 1 TeV. The green and blue dots show the TeV $\gamma$-ray data of VER J2016+371 observed by VERITAS \citep{2014ApJ...788...78A, 2018ApJ...861..134A}.}
	\label{fig:sed}
\end{figure*}

With the two-point source model, the spectra of PsA and PsB can be well fitted by the power-law models, and the spectral indices of them are fitted to be 2.902$\pm$0.109 and 1.689$\pm$0.232, respectively.
We also test the spectral curvature of PsA or PsB by adopting a log-parabola model,
while the fitting results do not improve significantly.
The integral photon flux of PsA and PsB in the energy range from 1 GeV to 1 TeV are $(2.69\pm0.32)\times10^{-9}$ and $(2.21\pm1.46)\times10^{-10}$ photon cm$^{-2}$ s$^{-1}$, respectively.

To derive the $\gamma$-ray spectral energy distributions (SEDs) of the two sources, 
we divide the data into eight logarithmically equal energy bins from 1 GeV to 1 TeV and repeat the likelihood fitting for each energy bin.
For the likelihood analysis, only the spectral normalizations of sources within $5^\circ$ from the center of ROI are left free, 
together with the normalizations of the two diffuse backgrounds.
And the spectral indices of these sources are fixed to be the best-fit values in the global likelihood analysis.
For the energy bin with TS value of PsA/PsB smaller than 5.0, an upper limit with 95\% confidence level is calculated.
The SEDs of PsA and PsB are shown in Figure \ref{fig:sed}, together with the global fitting results of the power-law models.

\section{Discussion}
\label{discussion}
  
\begin{figure*}[!htb]
	\centering
    \includegraphics[width=0.8\textwidth]{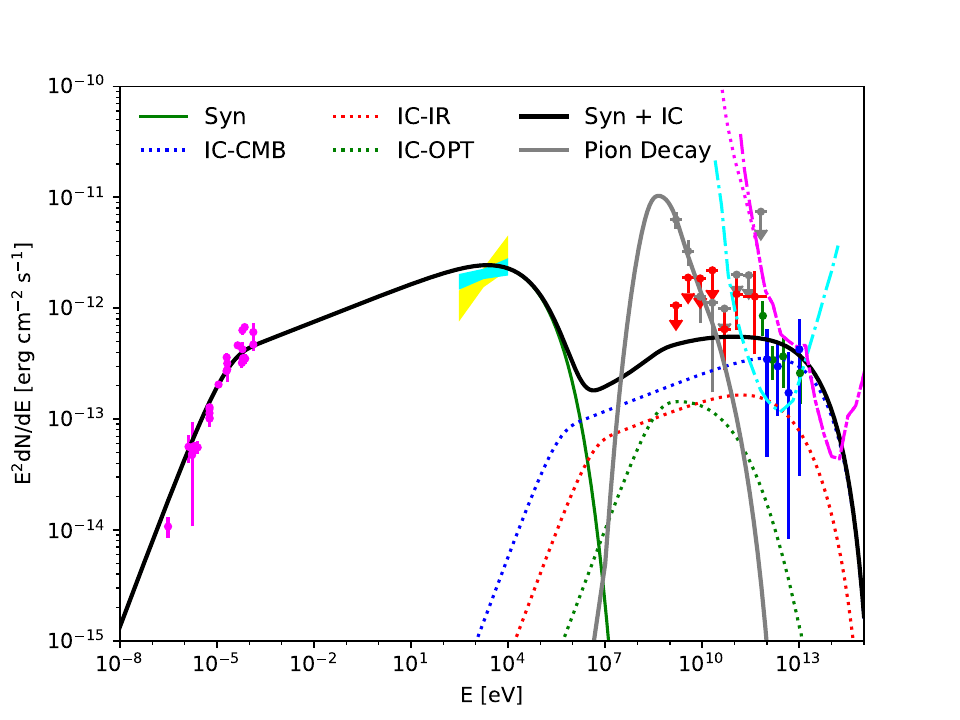}
	\caption{The multi-wavelength SED of VER J2016+371 with the leptonic model and PsA with hadronic model. The radio data marked by the magenta dots are from \citet{1975AJ.....80..437D, 1975A&A....38..461D, 1978A&A....70..389W, 1980A&A....84..237G, 1987A&AS...69..533M, 1990MNRAS.246..169P, 1991A&A...241..551W, 2006A&A...457.1081K, 2009MNRAS.396..365H, 2020MNRAS.496..723K,  2011A&A...536A..83S, 2022A&A...668A..39R, 2023ARep...67..963I}. The yellow and cyan butterflies represent the X-ray spectra of CTB 87 observed by {\em Chandra}\citep{2013ApJ...774...33M} and {\em XMM}-Newton \citep{2020MNRAS.491.3013G}. The TeV $\gamma$-ray data of VER J2016+371 are shown as the green and blue dots \citep{2014ApJ...788...78A, 2018ApJ...861..134A}. The green solid line shows the synchrotron component for VER J2016+371, and the dotted lines represent IC scattering with the different radiation fields. The black solid line is the sum of the different leptonic radiation components for VER J2016+371. The gray solid line indicates the hadronic model for PsA. The cyan and magenta lines represent the differential sensitivities of CTA-North \citep[50 hr;][]{2019scta.book.....C} and LHAASO (1 yr) with the different sizes of photomultiplier tube \citep[PMT;][]{2019arXiv190502773C}, respectively.}
	\label{fig:model}
\end{figure*}

The {\em Fermi}-LAT data analysis above shows that the $\gamma$-ray emission around CTB 87 could be separated into two components. 
PsA exhibits a soft GeV $\gamma$-ray spectrum that does not align with the TeV $\gamma$-ray spectrum of VER J2016+371.
While the hard GeV $\gamma$-ray spectrum of PsB could smoothly connect with the TeV $\gamma$-ray spectrum of VER J2016+371, thereby supporting PsB as the GeV counterpart of VER J2016+371.
The soft GeV $\gamma$-ray spectra typically observed in SNRs are usually associated with the interaction between the SNR shock and MCs, which is believed to be a hadronic process.
{\em Fermi}-LAT has detected many sources of such SNR-MC systems, like IC 443\citep{2013Sci...339..807A}, W44\citep{2025arXiv250103889A}, and W51C\citep{2016ApJ...816..100J}.
The typical $\gamma$-ray PWNe usually exhibit the hard GeV spectra, as seen in MSH 15-52 \citep{2010ApJ...714..927A}, HESS J1825–137\citep{2020A&A...640A..76P} and HESS J1356-645 \citep{2023ApJ...942..105L}, which are similar to the GeV spectrum of VER J2016+371.
Also compared with other $\gamma$-ray emitting PWNe \citep{2009ApJ...694...12M,2013ApJ...773...77A}, the central pulsar in CTB 87, PSR J2016+3711, with $\dot{\rm E}$ = 2.2 $\times$10$^{\rm 37}$ erg s$^{\rm -1}$ is energetic enough to power a $\gamma$-ray PWN.
Therefore, here we consider the scenario that the PWN powered by PSR J2016+3711 is favored for the $\gamma$-ray emission from VER J2016+371, and the soft $\gamma$-ray spectrum from PsA relates to the SNR-MC interaction.

For the $\gamma$-ray emission from PsA, a hadronic model is considered here.
The proton spectrum is modeled as a power-law with an exponential cutoff, given by the equation of $\rm dN/dE_p \propto E^{-\alpha_p} \exp(-E/E_{p,\rm cut})$, 
where $\rm \alpha_p$ and $\rm E_{p, \rm cut}$ are the spectral index and the cutoff energy of protons, respectively.
The distance of 6.1 kpc is adopted here \citep{2003ApJ...588..852K}.
\citet{2018ApJ...859..173L} gave a rough estimate of the density of the associated molecular gas at the eastern edge of the radio emission of CTB 87 with 34$-$43 cm$^{\rm -3}$. 
Here we adopt an average gas density of n$_{\rm gas} \sim$ 40 cm$^{\rm -3}$ in the hadronic model, and the fitted model for PsA is shown as the gray line in Figure \ref{fig:model}.
The soft $\gamma$-ray spectrum of PsA suggests a soft proton spectrum with an index of $\rm \alpha_p$ $\sim$ 2.9.
The cutoff energy of protons depicted in Figure \ref{fig:model} is set to be 1 TeV. 
However, this value is not well constrained and could potentially be significantly higher.
The total energy of protons above 1 GeV is estimated to be $\rm W_p \sim 5.0 \times 10^{49}(n_{\rm gas}/40\ {\rm cm}^{-3})^{-1}(d/6.1\ {\rm kpc})^2$ erg.
Such energy content of protons is comparable to that of the typical SNR-MC systems, which are explained by hadronic processes, such as W44 and W51C \citep{2019ApJ...874...50Z}.
The detection of MCs in the region of PsA also support the potential origin of the $\gamma$-ray emission \citep{2018ApJ...859..173L}.
The non-detection of the SNR shock suggests that it may have already dissipated into the ambient gas. 
Deeper observations, especially in the radio and X-ray bands, would be beneficial in revealing the structure of the shock.

We then consider the PWN scenario associated with PSR J2016+3711 for the $\gamma$-ray emission of VER J2016+371.
The spin-down luminosity of PSR J2016+3711 is $\dot{E} = 2.2 \times 10^{37} \ \rm erg \ s^{-1}$ with its characteristic age of $\tau_c = 11.1 \rm \ kyr$  \citep{2024MNRAS.528.6761L}.
This is similar to the properties of PSR B0833-45, which is associated with the $\gamma$-ray PWN Vela-X and has a spin-down luminosity of $\dot{E} = 6.9 \times 10^{36} \ \rm erg \ s^{-1}$ and a characteristic age of $\tau_c = 11.3 \ \rm kyr$ \citep{2001ApJ...556..380H}.
%The GeV $\gamma$-ray luminosity of VER J2016+371 in the energy range from 1 GeV to 1 TeV is calculated to be $L_{\rm 1 GeV - 1 TeV} \approx 2.7\times 10^{34}~(d/{\rm 6.1~kpc})^2~\rm{erg}~\rm{s}^{-1}$.
%Such value is several times lower than that of HESS J1420-607, whose GeV $\gamma$-ray luminosity is calculated to be $L_{\rm 1 GeV - 1 TeV} \approx 1.7\times 10^{35}~(d/{\rm 5.6~kpc})^2~\rm{erg}~\rm{s}^{-1}$.
The TeV $\gamma$-ray luminosity of VER J2016+371 in the energy range of 1-10 TeV is $L_{\rm 1-10 TeV} \approx 2.8\times 10^{33}~(d/{\rm 6.1~kpc})^2~\rm{erg}~\rm{s}^{-1}$.
Such value is also similar to that of Vela-X, whose TeV $\gamma$-ray luminosity is $L_{\rm 1-10 TeV} \approx 0.8\times 10^{33}~(d/{\rm 0.28~kpc})^2~\rm{erg}~\rm{s}^{-1}$.
%The X-ray luminosity in the range of 2-10 keV and the TeV $\gamma$-ray luminosity in the rang of 10-30 TeV are calculated to be  $L_{\rm 2-10 keV} \approx 9.6\times 10^{33}~(d/{\rm 6.1~kpc})^2~\rm{erg}~\rm{s}^{-1}$ \citep{2013ApJ...774...33M} and  $L_{\rm 10-30 TeV} \approx 1.1\times 10^{33}~(d/{\rm 6.1~kpc})^2~\rm{erg}~\rm{s}^{-1}$ \citep{2018ApJ...861..134A}, respectively. 
%the GeV $\gamma$-ray luminosity in the rang of 10-316 GeV is  $L_{\rm 10-316 GeV} \approx 1.4\times 10^{34}~(d/{\rm 6.1~kpc})^2~\rm{erg}~\rm{s}^{-1}$
Taking into account the TeV $\gamma$-ray luminosity of VER J2016+371 and the spin-down luminosity of PSR J2016+3711, the efficiency of converting the pulsar's rotational energy into $\gamma$-rays is calculated as $L_{\rm 1-10 TeV}$/$\dot{E}$ $\simeq$ 0.01\%.
This value is consistent with the efficiencies ( $\leq$ 10\%) observed in other TeV sources that have been identified as PWNe \citep{2013uean.book..359K,2018A&A...612A...2H}.

In PWN, the emission spanning from radio to X-ray bands is generally attributed to the synchrotron process, 
and the $\gamma$-ray emission is explained by the IC scattering process, which is a leptonic process.
Here, we adopt a simple one-zone leptonic model to explain the multi-wavelength data of VER J2016+371.
In the modeling, the electron distribution is assumed to be a broken power law spectrum with an exponential cutoff in the form of
$\frac{dN_e}{dE} \propto  \frac{(E/E_{br})^{-\gamma_1}}{1+(E/E_{br})^{\gamma_2-\gamma_1}}\rm{exp} \left(-\frac{E}{E_{e,cut}} \right)$ \citep{2011MNRAS.410..381B},
where $\gamma_1$ and $\gamma_2$ are spectral indices below and above the break energy $E_{\rm br}$, and $E_{\rm e,cut}$ is the cutoff energy of electron spectrum.
For the IC process, the infrared (IR) and optical (OPT) radiation fields are also considered except the cosmic microwave background (CMB).
With the the empirical relation introduced in \citet{2011ApJ...727...38S},  the temperatures and energy densities of the IR and OPT components are estimated to be about (T$_{\rm IR}$ = 30 K, u$_{\rm IR}$ = 0.2 eV cm$^{-3}$) and (T$_{\rm OPT}$ = 3000 K, u$_{\rm OPT}$ = 0.5 eV cm$^{-3}$), respectively. 
The radio observations cover a frequency range from 74 MHz to 32 GHz \citep{1975AJ.....80..437D, 1975A&A....38..461D, 1978A&A....70..389W, 1980A&A....84..237G, 1987A&AS...69..533M, 1990MNRAS.246..169P, 1991A&A...241..551W, 2006A&A...457.1081K, 2009MNRAS.396..365H, 2020MNRAS.496..723K,  2011A&A...536A..83S, 2022A&A...668A..39R, 2023ARep...67..963I}.
The X-ray observations of CTB 87 from {\em Chandra} \citep{2013ApJ...774...33M} and {\em XMM}-Newton \citep{2020MNRAS.491.3013G}, together with the TeV $\gamma$-ray spectrum of VER J2016+371 \citep{2014ApJ...788...78A, 2018ApJ...861..134A}, are also considered here.
The different radiation components are computed using {\em naima} package \citep{2015ICRC...34..922Z}, and the model fitting is shown in Figure \ref{fig:model}.

For the leptonic model, the break energy of electrons is fitted to be about 9 GeV to describe the spectral steepening in the high-frequency radio band. 
The spectral index below the break is fitted to be about 1.5 to model the low-frequency radio data. 
The high-frequency radio data and the hard X-ray spectrum constrain the spectral index of electrons above the break energy to be about 2.8, and the cutoff energy of electrons needs to be higher than $\sim $400 TeV.
It should be noted that such an electron spectrum would be expected to produce a $\gamma$-ray spectrum with an index of around 2.0 in the energy band of {\em Fermi}-LAT. 
However, this is still consistent with the hard GeV $\gamma$-ray spectrum of PsB, given its large statistical error.
A magnetic field strength of $\sim$ 7 $\mu$G and a total energy of electrons above 1 GeV of 7.8 $\times$ 10$^{\rm 48}$ erg are required to account for the observed flux in the radio and X-ray bands.
Such a value of the magnetic field strength is consistent with the typical values for $\gamma$-ray PWNe \citep{2014JHEAp...1...31T,2018A&A...609A.110Z,2024RAA....24g5016L}. 
With the values of magnetic field strength and the break energy in the leptonic model, the estimated synchrotron cooling timescale is significantly larger than the characteristic age of PSR J2016+3711 (11.1 kyr). 
This suggests that the spectral break is likely intrinsic to the electron spectrum injected into the PWN, rather than being caused by radiative losses \citep{2008ApJ...678L.113D}.
The majority of the pulsar's rotational energy budget can be roughly estimated using the formula $\rm E_{\rm rot}$ = $\rm \dot{\rm E_0} \dot{\rm \tau_0} (n-1)/2$ \citep{2018A&A...612A...2H}. Assuming a braking index of $\rm n = 3.0$ and an initial spin-down timescale with a typical value of $\dot{\rm \tau_0} = 1$ kyr, $\rm E_{\rm rot}$ is estimated to be approximately $\rm 8.4 \times 10^{\rm 49}$ erg, which suggest an efficiency of converting spin-down energy into relativistic electrons of $\sim 9\%$ by compared to the value of $\rm W_{\rm e}$ in the model.

\section{Summary}
\label{summary}
In this work, we studied the GeV $\gamma$-ray emission around CTB 87 using more than 16 years of {\em Fermi}-LAT data.
By using the observational data in the different energy ranges, we found two point sources with the different $\gamma$-ray spectra in the energy range of 1 GeV - 1 TeV.
PsA exhibits a very soft GeV $\gamma$-ray spectrum, which can be characterized by a power-law model with an index of $\sim$2.9. 
In contrast, PsB, which has a hard GeV $\gamma$-ray spectrum, is described by a power-law model with an index of $\sim$ 1.7.
For $\gamma$-ray emissions in SNRs, a soft GeV spectrum is typically associated with interactions between the SNR shock and MCs, as seen in the GeV spectra of SNRs like IC 443 and W44. 
Such $\gamma$-ray emissions are believed to result from hadronic processes.
Given the molecular environment surrounding CTB 87, which is associated with PsA, we propose that the soft $\gamma$-ray emission from PsA is related to the interaction between the SNR and MCs.
With a single power-law model for protons, the GeV spectrum of PsA could be explained with the soft proton spectrum. The total energy of protons is also comparable to that of the typical $\gamma$-ray SNR-MC systems.
The hard GeV $\gamma$-ray spectrum of PsB could smoothly connect with the TeV $\gamma$-ray spectrum of VER J2016+371, which supports the identification of PsB as the GeV counterpart of VER J2016+371.
Considering the PWN observations in the radio and X-ray bands, along with the typical hard GeV spectrum characteristic of PWNe, 
we suggest that VER J2016+371 originates from the PWN associated with PSR J2016+3711.
A leptonic model featuring a broken power-law spectrum for electrons can account for the multi-wavelength data of VER J2016+371. Moreover, the fitted values for the magnetic field strength and the energy content of electrons align with those typically observed in $\gamma$-ray PWNe.

CTB 87 presents an excellent target for studying the composite $\gamma$-ray SNRs. 
Deeper observations in the radio and X-ray bands to search for the SNR shock would be helpful in investigating the origin of the $\gamma$-ray emission in this region.
Meanwhile, the future detailed observations by LHAASO and CTA-North are also crucial to ultimately understanding the nature of CTB 87.

%% Also note that the akcnowlodgment environment does not support long amounts of text. If you have a lot of people and institutions to acknowledge, do not use this command. Instead, create a new \section{Acknowledgments}.
\begin{acknowledgments}
We would like to thank the anonymous referee for very helpful comments, which help to improve the paper. 
The authors acknowledge Qiancheng Liu for providing the observational image of molecular clouds for CTB 87 that appeared in \citet{2018ApJ...859..173L}.
This work is based on observations made with NASA's Fermi $\gamma$-ray Space Telescope, 
and supported by the Natural Science Foundation of Sichuan Province of China under grant No. 2024NSFSC0452, 
the Fundamental Research Funds for the Central Universities under grant No. 2682024CG002,
and the National Natural Science Foundation of China under grant No. 12103040.
\end{acknowledgments}

\bibliography{sample631}{}
\bibliographystyle{aasjournal}



\end{document}